\documentclass[12pt]{article}
\usepackage{geometry}                
\usepackage{graphicx}
\usepackage{amssymb}
\usepackage{epstopdf}
\usepackage{url}
\usepackage{hyperref}
\usepackage{fancyvrb}
\usepackage{color}
\usepackage{subfig}

\newcommand{\barrett}[0]{\texttt{barrett}}
\newcommand{\multinest}[0]{\texttt{MultiNest}}

\title{\sc Barrett: out-of-core processing of MultiNest output}
\author{Sebastian Liem\\
\small \textit{GRAPPA, University of Amsterdam, Netherlands}}%
\date{}

\begin{document}
\maketitle

\begin{abstract}
\barrett~is a Python package for processing and visualising statistical inferences made using the nested sampling algorithm \multinest.
The main differential feature from competitors are full out-of-core processing allowing \barrett~to handle arbitrarily large datasets.
This is achieved by using the HDF5 data format.
\end{abstract}

\multinest~\cite{Feroz:2007kg,Feroz:2008xx} is a nested sampling algorithm widely used to statistical inferences in various astrophysical and particle physics contexts.
The output produced by \multinest~are plain text files which in turn has to be processed in order to produce and visualise the desired marginalised posteriors, profile likelihood maps and sundry statistical results.
There are several packages available to process and visualise the \multinest~output~\cite{Fowlie:2016hew, refId0, GetDist}.
They all, however, assumes that the size of the output to be processed is smaller than the available system memory. 
But, for large dimensional problem, especially when doing a profile likelihood analysis, the size of that output can become $\mathcal{O}(10-100\ \mathrm{GB})$ which violates that assumption.
This paper introduces \barrett~which is a Python package that can handle such large datasets.

Marginalisation and profiling are two ways of reducing the dimensionality of the posterior and likelihood respectively. 
This is done either to eliminate uninteresting, i.e. nuisance, parameters or to reduce the dimensionality to one or two dimension in order to be able to visualise the results. 
Denoting the parameters of interests as $\delta$, and the ones to be eliminated as $\xi$ the marginalised posterior and profile likelihood can be defined as:
\begin{eqnarray*}
p(\delta| \mathrm{data}) = \int p(\delta, \xi | \mathrm{data}) \, \mathrm{d}\xi \qquad L(\mathrm{data} | \delta) = \sup_{\xi} L(\mathrm{data} | \delta, \xi)
\end{eqnarray*}

where $p(\delta, \xi | \mathrm{data})$ and $L(\mathrm{data} | \delta, \xi)$ is the full posterior and likelihood respectively. 
In the context of the \multinest~output, the operation underlying both the marginalisation of posteriors and the profiling of likelihoods is binning.
The output is a list of points in the scanned parameter space with an associated posterior weight and likelihood.
To profile or marginalise one define bins in the parameters of interest, and group every point into these bins.
The marginalised posterior in each bin then is the sum of the posterior weights of all the points in each bin.
The profile likelihood in each bin is the largest likelihood of all the points in each bin.

\barrett~make use of the fact that binning is an operation that can be done in parts. 
The \multinest~output is divided into parts small enough to fit system memory.
Each part is then read into memory in turn and binned.
The partial results can easily be combined by either summing them (marginalisation) or taking the bin maxima (profiling).

Besides the usual components of scientific Python (\texttt{numpy},  \texttt{scipy}, and  \texttt{mat\-plot\-lib}~\cite{numpy, scipy, matplotlib}) the only dependency is the \texttt{h5py} package~\cite{h5py} which provides the Python interface to the HDF5~\cite{hdf5} data format.
\barrett~uses HDF5 files because it offers two important advantages over the plain text file produced by \multinest.
First, HDF5 is a binary format which takes much less disk space when compared to plain text.
Secondly, the HDF5 data format allows one to 'chunk' the data, i.e. divides it into parts while still storing it in the same file.
The library then only loads the chunks accessed into memory.
While the first advantage is a nice bonus it is the second that is essential for \barrett~allowing it to perform the out-of-core binning described in the paragraph above.

\section{Installation}
\barrett~is available on PyPI and can therefor be installed as easily as 
\begin{Verbatim}[frame=leftline,framesep=2ex,framerule=0.8pt,fontsize=\small]
$ pip install barrett
\end{Verbatim}

Alternately the code is available at \url{https://github.com/sliem/barrett} where any issues or bugs can also be reported. 

\section{Description}

\barrett~has four submodules; \texttt{data}, \texttt{posterior}, \texttt{profilelikelihood}, and \texttt{util}.

\begin{description}
\item[\texttt{barrett.data}] implements methods for modifying data (e.g. log, change units) or calculate depended variables (e.g. mean squark mass) in an out-of-core fashion.

\item[\texttt{barrett.posterior}] defines two classes \texttt{oneD} and \texttt{twoD} for 1D and 2D marginalised posteriors respectively. Each class performs the marginalisation on initialisation, and implements a plotting method to visualise the results. The \texttt{twoD} class also implements a method for calculating the credibility regions.

\item[\texttt{barrett.profilelikelihood}] also defines two classes named \texttt{oneD} and \texttt{twoD} but for the 1D and 2D profile likelihoods. The profiling is done on instance initialisation, and the classes define plotting methods. The \texttt{twoD} class can also calculate the desired confidence regions.

\item[\texttt{barrett.util}] contains, as the name suggest, functions of varying utility. Perhaps, the most useful one is \texttt{convert\_chain()} which converts from the plain text output of \multinest~to the HDF5 format used by \barrett.

\end{description}

The default \texttt{plot()} methods are functional but, by design, very basic.
Rather than putting another layer of API on top of \texttt{matplotlib} \barrett~encourage the user to directly work with \texttt{matplotlib}.
This can be done either by directly accessing the variables in the classes or by subclassing them and override the \texttt{plot()} method.

\section{Example}

\begin{figure}[htbp!]
\centering
\subfloat[][default plot method]{
\includegraphics[width=0.45\linewidth,page=1]{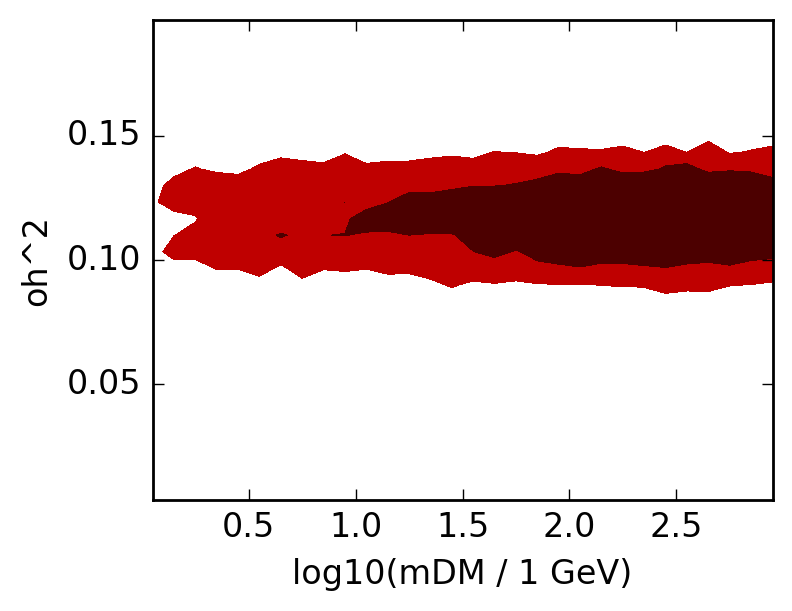}
\label{fig:ex1}
}
\subfloat[][custom plot method]{
\includegraphics[width=0.45\linewidth,page=1]{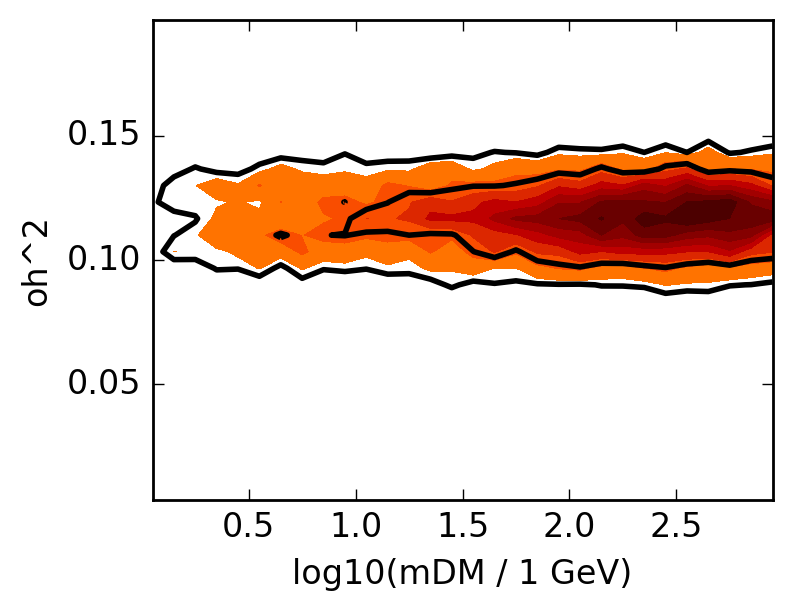}
\label{fig:ex2}
}
\caption{The resulting plots of the marginalised posterior using the two example code snippets. }
\label{fig:examples}
\end{figure}

In the following two examples \barrett~will be use to marginalise a posterior two dimensions and then visualise it.
The details of the physics isn't of interest, but the parameters used are the dark matter mass and relic density. 
The results of the example codes are in Fig. \ref{fig:examples}.
The first example shows the basic usage of \barrett.

\begin{Verbatim}[frame=leftline,framesep=2ex,framerule=0.8pt,fontsize=\small]
import barrett.posterior as posterior
import matplotlib.pyplot as plt

h5file = 'RD.h5'
xvar = 'log(m_{\chi})'
yvar = 'Omega_{\chi}h^2'
P = posterior.twoD(h5file, xvar, yvar, 
                   xlimits=(0,3), ylimits=(0.0, 0.2),
                   xbins=30, ybins=30)

fig, ax = plt.subplots(ncols=1, nrows=1)
P.plot(ax)

ax.set_xlabel('log10(mDM / 1 GeV)')
ax.set_ylabel('oh^2')
fig.savefig('ex1.png', dpi=200, bbox_inches='tight')
\end{Verbatim}

The second example shows how to customise the plot by subclassing and overriding the \texttt{plot()} method.
\newpage
\begin{Verbatim}[frame=leftline,framesep=2ex,framerule=0.8pt,fontsize=\small]
class myTwoD(posterior.twoD):
    def plot(self, xlabel, ylabel, fig_path):
        fig, ax = plt.subplots(ncols=1, nrows=1)

        # plot heatmap of the posterior.
        posterior.twod.plot(self, ax, levels=none)

        # plot one and two sigma contours.
        x, y = np.meshgrid(self.xcenters, self.ycenters)
        levels = np.append(self.credibleregions([0.95, 0.68]),
        	           self.pdf.max())
        ax.contour(x, y, self.pdf, levels=levels, colors='k')

        ax.set_xlabel(xlabel)
        ax.set_ylabel(ylabel)
        fig.savefig(fig_path, dpi=200, bbox_inches='tight')

P = myTwoD(h5file, xvar, yvar, 
           xlimits=(0,3), ylimits=(0.0, 0.2),
           xbins=30, ybins=30)
P.plot('log10(mDM / 1 GeV)', 'oh^2', 'ex2.png')
\end{Verbatim}

Additional examples can be found on \url{https://github.com/sliem/barrett}.

\newpage
\bibliographystyle{JHEP}
\bibliography{Barrett}

\providecommand{\href}[2]{#2}\begingroup\raggedright\begin{thebibliography}{10}

\bibitem{Feroz:2007kg}
F.~Feroz and M.~P. Hobson, {\it {Multimodal nested sampling: an efficient and
  robust alternative to MCMC methods for astronomical data analysis}},  {\em
  Mon. Not. Roy. Astron. Soc.} {\bf 384} (2008) 449,
  [\href{http://xxx.lanl.gov/abs/0704.3704}{{\tt arXiv:0704.3704}}].

\bibitem{Feroz:2008xx}
F.~Feroz, M.~P. Hobson, and M.~Bridges, {\it {MultiNest: an efficient and
  robust Bayesian inference tool for cosmology and particle physics}},  {\em
  Mon. Not. Roy. Astron. Soc.} {\bf 398} (2009) 1601--1614,
  [\href{http://xxx.lanl.gov/abs/0809.3437}{{\tt arXiv:0809.3437}}].

\bibitem{Fowlie:2016hew}
A.~Fowlie and M.~H. Bardsley, {\it {Superplot: a graphical interface for
  plotting and analysing MultiNest output}},
  \href{http://xxx.lanl.gov/abs/1603.00555}{{\tt arXiv:1603.00555}}.

\bibitem{refId0}
{Buchner, J.}, {Georgakakis, A.}, {Nandra, K.}, {Hsu, L.}, {Rangel, C.},
  {Brightman, M.}, {Merloni, A.}, {Salvato, M.}, {Donley, J.}, and {Kocevski,
  D.}, {\it X-ray spectral modelling of the agn obscuring region in the cdfs:
  Bayesian model selection and catalogue},  {\em A\&A} {\bf 564} (2014) A125.

\bibitem{GetDist}
A.~Lewis, {\it Getdist},  2015--.
\newblock \url{https://github.com/cmbant/getdist/} [Online; accessed
  2016-07-31].

\bibitem{numpy}
S.~van~der Walt, S.~C. Colbert, and G.~Varoquaux, {\it The numpy array: A
  structure for efficient numerical computation},  {\em Computing in Science
  Engineering} {\bf 13} (March, 2011) 22--30.

\bibitem{scipy}
E.~Jones, T.~Oliphant, P.~Peterson, et~al., {\it {SciPy}: Open source
  scientific tools for {Python}},  2001--.
\newblock \url{http://www.scipy.org/} [Online; accessed 2016-07-31].

\bibitem{matplotlib}
J.~D. Hunter, {\it Matplotlib: A 2d graphics environment},  {\em Computing in
  Science Engineering} {\bf 9} (May, 2007) 90--95.

\bibitem{h5py}
A.~Collette, {\it {HDF5 for Python}},  2008--.
\newblock \url{http://www.h5py.org/} [Online; accessed 2016-07-31].

\bibitem{hdf5}
{The HDF group}, {\it {Hierarchical Data Format, version 5}},  1997--.
\newblock \url{http://www.hdfgroup.org/HDF5/}, [Online; accessed 2016-07-31].

\end{thebibliography}\endgroup

\end{document}